\documentclass[prb,twocolumn,amsmath,amssymb]{revtex4}
\usepackage{graphicx}
\usepackage{dcolumn}
\usepackage{bm}

\begin{document}


\title{Determination of the gate-tunable bandgap and tight-binding parameters
in bilayer graphene using infrared spectroscopy}

\author{A. B. Kuzmenko}
\affiliation{D\'epartement de Physique de la Mati\`ere
Condens\'ee, Universit\'e de Gen\`eve, CH-1211 Gen\`eve 4,
Switzerland}

\author{I. Crassee}
\affiliation{D\'epartement de Physique de la Mati\`ere
Condens\'ee, Universit\'e de Gen\`eve, CH-1211 Gen\`eve 4,
Switzerland}

\author{D. van der Marel}
\affiliation{D\'epartement de Physique de la Mati\`ere
Condens\'ee, Universit\'e de Gen\`eve, CH-1211 Gen\`eve 4,
Switzerland}

\author{P. Blake}
\affiliation{Manchester Centre for Mesoscience and
Nanotechnology, University of Manchester, Manchester M13 9PL,
UK }

\author{K. S. Novoselov}
\affiliation{Manchester Centre for Mesoscience and
Nanotechnology, University of Manchester, Manchester M13 9PL,
UK }

\date{\today}

\begin{abstract}
We present a compelling evidence for the opening of a bandgap
in exfoliated bottom-gated bilayer graphene by fitting the
gate-voltage modulated infrared reflectivity spectra in a large
range of doping levels with a tight-binding model and the Kubo
formula. A close quantitative agreement between the
experimental and calculated spectra is achieved, allowing us to
determine self-consistently the full set of
Slonczewski-Weiss-McClure tight-binding parameters together
with the gate-voltage dependent bandgap. The doping dependence
of the bandgap shows a good agreement with the existing
calculations that take the effects of self-screening into
account. We also identify certain mismatches between the
tight-binding model and the data, which can be related to
electron-electron and electron-phonon interactions.
\end{abstract}

\pacs{}
\maketitle

\section{Introduction}

Bilayer graphene has recently attracted much attention
motivated by a broad spectrum of unusual electronic properties
and a number of possibilities for applications. It represents
the simplest system, where the effects caused by a coupling
between graphene layers can be studied and exploited. Although
the interlayer coupling is much weaker than the in-plane
chemical bonding, it results in profound differences between
electronic and transport properties of monolayer and bilayer
graphene, as exemplified by the anomalous quantum Hall effect
\cite{NovoselovNature05,ZhangNature05,NovoselovNP06}. Another
notable dissimilarity is related to the behavior of these
systems in a perpendicular electric field. While in zero field
both of them are zero-gap semiconductors (or zero-overlap
semimetals), in bilayer graphene a bandgap is generated in the
presence of the field, due to an introduced asymmetry of the
electrostatic potential on the two planes
\cite{McCannFalkoPRL06,NilssonPRB07,McCannPRB06,CastroPRL07}.
Importantly, the bandgap can be tuned continuously, either by
applying gate voltage \cite{OostingaNatMat08} or chemically
\cite{OhtaScience06}, which, in combination with a high
mobility of charge carriers, opens new unexplored avenues for
using bilayer graphene in field effect transistors (FETs)
\cite{RyzhiiAPL07,LinNanoLett09,FioriIannacconiIEEE09} and
other electronic devices.

Using angle-resolved photoemission spectroscopy (ARPES), Ohta
et al.\cite{OhtaScience06} indeed observed a bandgap in bilayer
graphene epitaxially grown on top of silicon carbide and doped
chemically with potassium. A more "clean" way of introducing
charge carriers is by applying electric field using gate
electrodes. This technique can be most easily applied to
exfoliated samples, produced by micromechanical cleavage of
graphite \cite{NovoselovScience04}. Apart from the bandgap
generation, applying a gate voltage has also the usual doping
effect. In order to control the doping and the bandgap
independently, Oostinga et al.\cite{OostingaNatMat08}
fabricated two electrodes on both sides of the sample and found
an insulating state, when gate voltages of opposite sign were
applied to the electrodes. This showed the existence of the
bandgap, although the determination of its exact value from the
DC measurements was not possible.

Infrared spectroscopy, which is one of the most direct methods
to measure the bandgap and other band characteristics in
conventional semiconductors, is clearly technique of choice
also in the case of bilayer graphene. The two-dimensionality of
this material perfectly matches geometrical requirements of an
optical experiment. Moreover, the possibility of changing the
chemical potential and the bandgap with the gate voltage
supplies an unprecedented amount of additional information
\cite{LiNP08,JiangPRL07,HenriksenPRL08,WangScience08,ZhangPRB08,LiPRL09,KuzmenkoPRB09,MakCM09,ZhangNature09}
compared to standard optical measurements.

The tight-binding theory is widely used to describe the low
energy $\pi$ bands in graphitic materials. In the case of
graphite, a set of tight-binding parameters, known as the
Slonczewski-Weiss-McClure (SWMcC) model
\cite{McClurePR57,SlonczewskiWeissPR58}, was very successful in
describing quantitatively the de Haas-van Alphen effect and
optical spectra \cite{DresselhausAdvPhys81}. Therefore we one
can expect that it will also apply to bilayer graphene, if the
bandgap is properly included. It appears that all SWMcC
parameters influence the optical conductivity for photon
energies below 1 eV. However, the effects of different
parameters are rather dissimilar and not all of them can be
easily extracted from the spectra.

Several calculations of the optical conductivity of bilayer
graphene within the tight binding approach were done. Nilsson
et al.\cite{NilssonPRL06} and Abergel and Falko
\cite{AbergelFalkoPRB07} considered the simplest model, which
contains only the nearest-neighbor in-plane and interplane
hopping terms ($\gamma_{0} \sim 3$ eV and $\gamma_{1} \sim 0.4$
eV respectively) and found that the optical conductivity is
marked by a profound structure at the photon energy $\hbar
\omega \sim \gamma_{1}$ (we shall refer to this structure as
the $\gamma_{1}$ - peak). Nicol and Carbotte
\cite{NicolCarbottePRB08} extended this model to include a
bandgap and finite doping. Zhang et al. \cite{ZhangPRB08}
studied the role of additional parameters responsible for the
electron-hole asymmetry.

Using infrared techniques, Wang et al \cite{WangScience08}
observed, in agreement with aforementioned predictions, a
profound anomaly in the spectra of gated bilayer graphene at
$\hbar\omega \sim \gamma_{1}$. Later on, a clear electron-hole
asymmetry was found in the infrared spectra by Li et al
\cite{LiPRL09} and by Kuzmenko et al \cite{KuzmenkoPRB09},
which allowed determining more SWMcC parameters. However, the
presence of the bandgap in these studies, although expected,
was not evident. In Ref.\onlinecite{LiPRL09} no experimental
signatures of the bandgap were reported and in
Ref.\onlinecite{KuzmenkoPRB09} only a partial agreement between
the experimental data and a tight-binding calculation including
a bandgap obtained theoretically \cite{CastroPRL07} was found.
As it will be discussed below, this is explained by the fact
that the manifestation of the bandgap in doped graphene is more
subtle, and therefore requiring more accurate optical
measurements and delicate analysis than in the case of an
undoped sample. In the latter case a sharp absorption threshold
corresponding to the electron-hole excitations across the gap
is expected.

Such a structure was indeed observed by Zhang et al.
\cite{ZhangNature09}, who measured infrared absorption of
double gated bilayer graphene, where the electric field and
doping could be controlled independently
\cite{OostingaNatMat08}. A bandgap up to 250 meV was observed
in an undoped sample in the presence of the largest applied
fields, which is a rather promising sign for the use of bilayer
graphene in electronics. Although for the large values of the
bandgap the match between the experiment and a tight-binding
calculation was very close, at low gate voltages, where the
absorption threshold was beyond the experimentally accessible
range, the quantitative agreement between the experimental and
theoretical curves turned out to be rather poor.

In general, the quantitative agreement between infrared spectra
of bilayer graphene and tight-binding model was up to now not
very good. Apart from the mentioned discrepancies, the measured
height of the $\gamma_{1}$ peak in Ref.\onlinecite{ZhangPRB08}
was about two times larger than the calculated value. Mak et
al. reported the opening of a bandgap in top-gated bilayer
graphene \cite{MakCM09}, where a small thickness of the gate
insulating layer allowed very efficient doping. The
quantitative agreement with the tight-binding theory, however,
was limited, which the authors attributed to many-body
correlation effects.

A common problem that one encounters when analyzing infrared
spectra of graphene is their sensitivity to several band
parameters, including the bandgap, making their separate
extraction quite complicated. Another issue is a possible
inhomogeneity of the doping level, which has a similar effect
on the optical spectra as an elevated temperature, as will be
shown below. In the double-gate experiments, the deposition of
the top gate on top of graphene may affect the band structure
and increase electronic scattering, not to mention a more
complicated optical multilayer model that has to be used in
order to extract the optical conductivity of graphene.
Therefore, in the present study we fit directly the measured
reflectivity spectra of bottom gated bilayer graphene with a
tight-binding model that involves the bandgap, the SWMcC
parameters, scattering rate, temperature and the impurity
doping as adjustable parameters. We find that a good
quantitative agreement can in fact be achieved, which implies
that the band structure of bilayer graphene is well captured by
the tight binding model. Nevertheless, certain discrepancies
remain that may eventually be related to many-body effects.

The remaining sections are organized as follows. In section
\ref{SectionTechniques} we describe the sample preparation,
infrared experiment and a specially developed technique of
direct fitting of the whole set of reflectivity spectra with a
tight-binding model and the Kubo formula. In section
\ref{SectionResults}, the results of the infrared measurements,
their fits, and the doping dependence of the bandgap and the
chemical potential are shown. In section
\ref{SectionDiscussion} we discuss the reliability of
extracting the bandgap and compare its gate-voltage dependence
with the existing theoretical predictions. We demonstrate a
practical way of visualizing important features of the band
structures using the experimental reflectivity data. We also
discuss the advantages and limitations of the tight-binding
model in describing infrared spectra.

\section{Techniques}\label{SectionTechniques}

\subsection{Sample}\label{SubsectionTechniquesSample}

A relatively large ($\sim$ 100 $\mu$m) flake of bilayer
graphene produced by micromechanical cleavage
\cite{NovoselovScience04} of graphite single crystals on top of
a SiO$_{2}$ (300 nm)/n-Si substrate was chosen for infrared
experiments. Lithographically deposited leads to the flake and
the gate (doped silicon) allowed a simultaneous measurement of
the DC conductivity $\sigma_{DC}$ and the infrared reflectivity
$R$ as functions of the gate voltage $V_{g}$. By sweeping
$V_{g}$ from the positive to the negative values, one
continuously varies the doping from electron to hole type. The
charge concentration can be determined using the relation $n$ =
$\alpha (V_{g} - V_{CN})$, where the coefficient $\alpha = 7.2
\times10^{10}$ cm$^{-2}/$V is given by the electric capacitance
of the oxide layer. Usually the charge neutrality point
$V_{CN}$ is related to the minimum $V_{min}$ of the
$\sigma_{DC}(V_{g})$ curve. However, from the fits of the
optical spectra described below we found that $V_{CN}$ (which
is about -22 V for the presented series of measurements) is
slightly different from $V_{min}$ ($\approx$ -29 V). It is
likely that this is related to the large flake dimensions and a
doping inhomogeneity near electrical contacts.

The negative value of $V_{CN}$ is due to charge transfer by
adsorbed gas molecules (environmental doping)
\cite{SchedinNM07}. In order to reduce this effect, the sample
was annealed in a H$_{2}$-N$_{2}$ atmosphere at 150 $^{\circ}$C
before each series of measurements. Although the annealing
indeed shifts $V_{CN}$ closer to zero bias, we found that, in
contrast to the case of monolayer graphene, it never results in
$V_{CN} = 0$. It is possible that the remaining dopants are
located either between the flake and the substrate or are even
intercalated between the carbon layers. As we will see below,
this correlates with the doping dependence of the bandgap.

\subsection{Optical experiment}\label{SubsectionTechniquesOptics}


\begin{figure}[thb]
   \includegraphics[width=8.5cm,clip=true]{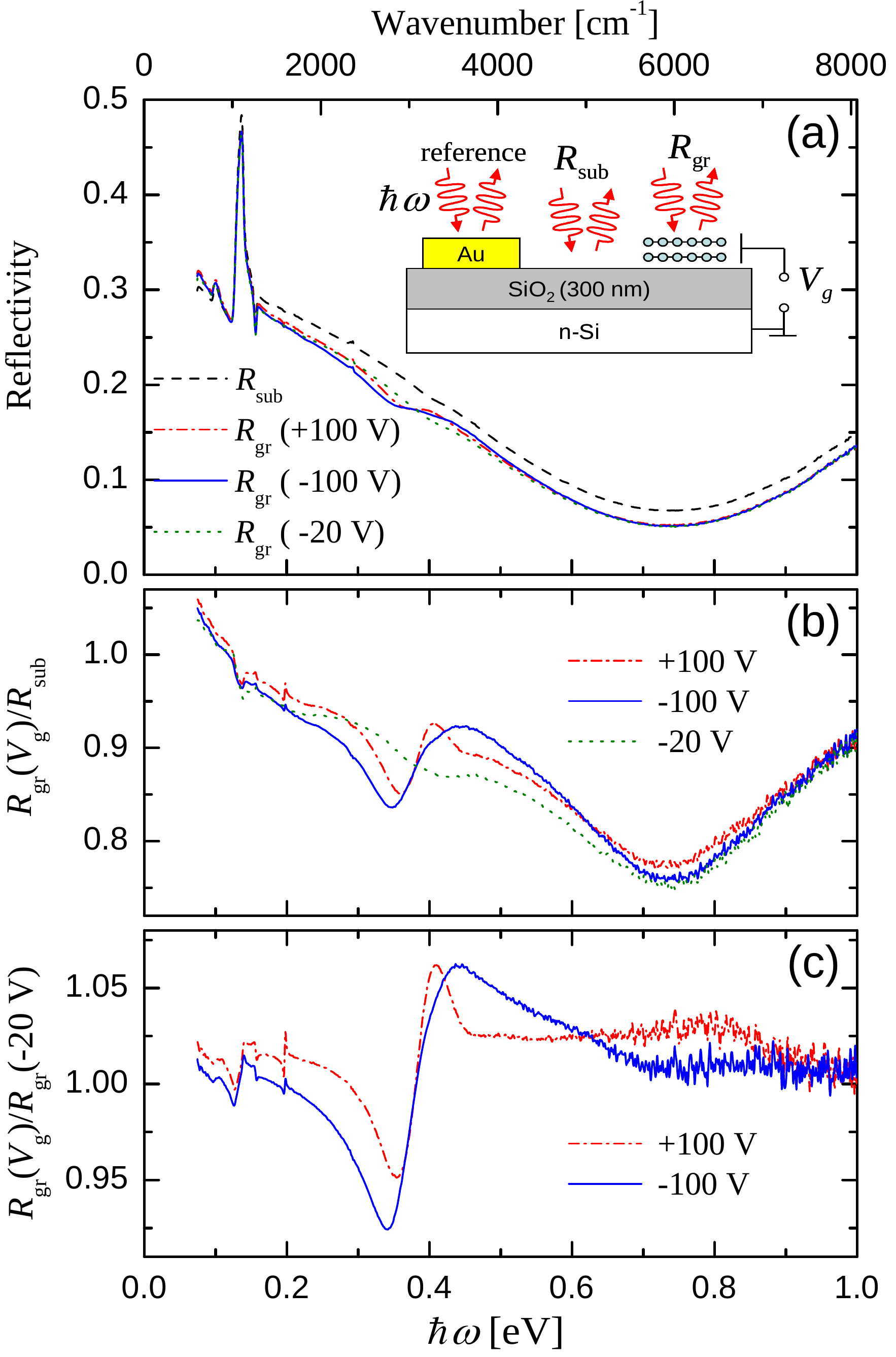}
   \caption{(Color online)
   (a) Absolute (gold-normalized) reflectivity of bare oxide and graphene at $V_{g}$ = +100, -20 and -100
   V). The substrate temperature is 10 K. (b) Reflectivity of
   graphene normalized to bare oxide. (c) Self-normalized reflectivity of graphene $R_{gr}(V_{g})/R_{gr}( -20 \mbox{ V})$.}
   \label{FigSpectraMeas}
\end{figure}

Optical reflectivity spectra in the range of photon energies
0.06 - 1 eV were collected at nearly normal incidence using an
IR microscope Bruker Hyperion 2000 attached to a Fourier
transform spectrometer with a standard globar source.
Appreciable signal could be obtained using the beam spot down to
10-15 $\mu$m. However, prior to each series of measurements we optimized 
in a try-and-error fashion the spot size in order to maximize the signal-to-noise ratio 
while avoiding unphysical spectral artefacts related to the finite sample dimensions 
and apparatus issues. The spectral resolution was 1 meV. 
The sample was mounted in a He flow cryostat; a specially made 
sample holder allowed us inserting the sample with a minor delay after the annealing.

Absolute reflectivity spectra of the bare substrate
$R_{sub}(\omega)$ and the flake $R_{gr}(\omega)$ were obtained
using a gold patch deposited near the sample as a reference
(Fig.\ref{FigSpectraMeas}a). The substrate and graphene spectra
look similar: they all show a prominent minimum at 0.75 eV due
to the Fabry-Perot effect in the SiO$_{2}$ layer and intense
peaks below 0.15 eV originating from dipole-active lattice
vibrations in silicon oxide. The change of reflectivity
introduced by graphene is better seen in
Fig.\ref{FigSpectraMeas}b, where $R_{gr}(\omega)$ is normalized
to $R_{sub}(\omega)$. One can see that graphene introduces a
significant infrared contrast, especially close to the
Fabry-Perot minimum. Notably, the same effect in the visible
range makes graphene detectable by human eye \cite{BlakeAPL07}.

From Fig.\ref{FigSpectraMeas}b it is also clear that varying
the gate voltage has a strong effect on reflectivity. To see it
even better, one can normalize $R_{gr}$ by its value at $V_{g}
= -20$ V, which is close to the charge neural point
(Fig.\ref{FigSpectraMeas}c). Note that the SiO$_{2}$ phonon
features do not fully cancel after the normalization, because
of a non-linear character of the contribution of the substrate
to the spectra. At 0.2 eV one can see a sharp structure related
to the infrared active phonon mode in graphene, which has a
Fano shape and strongly increases as a function of the gate
voltage due to a coupling to electronic interband transitions,
as discussed in detail in Ref. \onlinecite{KuzmenkoCM09}. As we
shall see below, the other changes are due to a combination of
the doping effect and the opening of the bandgap.

One should note that the diffraction of electromagnetic
radiation may affect the measured reflectivity at low energies,
where the wavelength becomes comparable to the spot size. Due
to the diffraction and other systematic uncertainties, the
absolute accuracy of $R_{gr}$ and $R_{gr}/R_{sub}$ is about
0.01-0.02 as we determined by varying the position and the size
of the beam spot on the sample. In contrast, the
self-normalized reflectivity $R_{gr}(V_{g})/R_{gr}(-20 \mbox{
V})$ can be measured much more accurately (with the uncertainty
less than 0.002) since it does not require any reference
measurement and therefore does not involve any mechanical
movements. In order to minimize the influence of weak drifts of
the signal, taking spectrum at each gate voltage was
immediately followed by a separate measurement at the charge
neutral point.

\subsection{Data modelling}\label{SubsectionTechniquesModelling}

For the data analysis we chose the SWMcC
\cite{McClurePR57,SlonczewskiWeissPR58} tight binding
description of the $\pi$ bands in bilayer graphene
\cite{CastroNetoRMP09}. We begin with considering it simply as
a band structure parametrization in order to extract
information about the electronic bands from the infrared data.
Later on we shall discuss the limitations of this description
as a physical model based on the quality of the obtained fits.

\begin{figure}[htp]
\includegraphics[width=7cm]{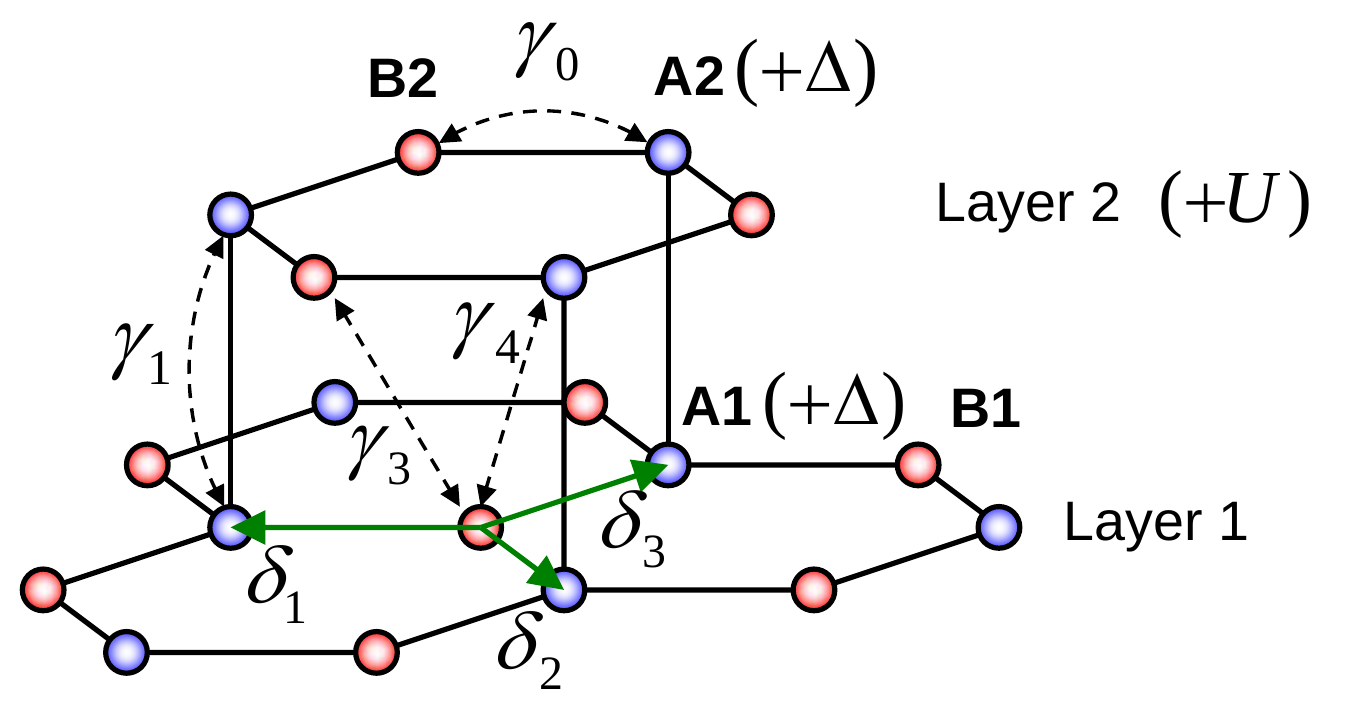}\\
\caption{(Color online) Crystal structure of Bernal stacked
bilayer graphene and the considered tight-binding parameters.}
\label{FigStructure}
\end{figure}

The structure of the Bernal-stacked bilayer graphene is shown
in Fig.\ref{FigStructure}. Each layer has two sublattices: A1
and B1 (bottom layer) and A2 and B2 (top layer). Atoms A1 and
A2 are on top of each other, while the atoms B1 and B2 are
shifted horizontally by the vectors $\vec{\delta}_{1,2,3}$
connecting nearest neighbors within one layer. The SWMcC
Hamiltonian involves the in-layer nearest-neighbor hopping
$\gamma_{0}$ and three interlayer hopping terms: $\gamma_{1}$
(between A1 and A2), $\gamma_{3}$ (between B1 and B2) and
$\gamma_{4}$ (between A1 and B2 or between B1 and A2). In
addition, the on-site energy difference $\Delta$ between the
positions A1 and B1 (A2 and B2) is introduced. Following the
standard procedure
\cite{McCannFalkoPRL06,LuPRB06,McCannPRB06,NilssonPRB07,CastroPRL07},
we add to the SWMcC model an extra parameter $U$ in order to
describe the difference between the (screened) electrostatic
potential of the top and the bottom layers in the external
field. Note that $U$ gives exactly the separation between the
electron and hole bands at the K point and is slightly larger
than the true bandgap $\Delta_{g}\approx |U|\gamma_{1}/(U^2 +
\gamma^2_{1})^{1/2}$. In the basis $|\mbox{B1}\rangle$,
$|\mbox{A1}\rangle$, $|\mbox{A2}\rangle$ and
$|\mbox{B2}\rangle$ the present Hamiltonian reads as follows:
\begin{equation}
H(\vec{q})=\left(
  \begin{array}{cccc}
    0 & \gamma_{0}\phi & -\gamma_{4}\phi & \gamma_{3}\phi^{*} \\
    \gamma_{0}\phi^{*} & \Delta & \gamma_{1} & -\gamma_{4}\phi \\
    -\gamma_{4}\phi^{*} & \gamma_{1} & \Delta + U & \gamma_{0}\phi \\
    \gamma_{3}\phi & -\gamma_{4}\phi^{*} & \gamma_{0}\phi^{*} & U \\
  \end{array}
\right)\label{EqHam}
\end{equation}

\noindent where $\phi = e^{i\vec{q}\vec{\delta}_{1}} +
e^{i\vec{q}\vec{\delta}_{2}} + e^{i\vec{q}\vec{\delta}_{3}}$,
$\vec{q}$ is the electronic momentum.

Within the linear response theory, the Kubo formula can be used
to calculate the complex optical conductivity $\sigma(\omega)$.
In the case of a thin layer, a physically more relevant
quantity is the optical sheet {\em conductance} $G(\omega) =
\sigma(\omega) d$ where $d$ is the layer thickness. The total
conductance consists of the Drude, the interband and the
high-frequency terms:
\begin{eqnarray}
G(\omega) = G_{\mbox{\tiny D}}(\omega) + G_{\mbox{\tiny
IB}}(\omega) + G_{\infty}(\omega).\label{EqG}
\end{eqnarray}
\noindent The first two terms can be obtained using the
expressions:

\begin{eqnarray}
G_{\mbox{\tiny D}}(\omega) =\frac{2G_{0}}{\pi^2}\sum_{i}\int
d^2\vec{q}\left|\left\langle \vec{q},i\left|\frac{\partial
H}{\partial q_x
}\right|\vec{q},i\right\rangle\right|^2\nonumber\\
\times\left(\frac{-\partial
f(\epsilon_{\vec{q},i})}{\partial\epsilon}\right)\frac{i}{\hbar\omega
+ i \Gamma_{\mbox{\tiny D}}}\label{EqGD}\\
G_{\mbox{\tiny IB}}(\omega) = \frac{2G_{0}}{\pi^2}\sum_{i,j\neq
i}\int d^2\vec{q}\left|\left\langle
\vec{q},i\left|\frac{\partial H}{\partial q_x
}\right|\vec{q},j\right\rangle\right|^2\nonumber\\
\times\frac{f(\epsilon_{\vec{q},i}) -
f(\epsilon_{\vec{q},j})}{\epsilon_{\vec{q},j} -
\epsilon_{\vec{q},i}} \frac{i}{\hbar\omega -
\epsilon_{\vec{q},j} + \epsilon_{\vec{q},i} + i \Gamma}
\label{EqGIB}
\end{eqnarray}

\noindent where $G_{0} = e^2/4\hbar \approx 6.08\times
10^{-5}\mbox{ }\Omega^{-1}$ is the universal AC conductance of
monolayer graphene
\cite{AndoJPSJ02,GusyninPRL06,FalkovskyVarlamovEPJB07,NairScience08}
and graphite \cite{KuzmenkoPRL08}, $\epsilon_{\vec{q},i}$ are
the electronic band energies, $\Gamma$ is the electronic
broadening parameter and $f(\epsilon) = \left\{1 +
\exp\frac{\epsilon - \mu}{k_{B}T}\right\}^{-1}$ is the
Fermi-Dirac distribution. For the derivation of these formulas
we can refer, for example, to Ref.
\onlinecite{FalkovskyVarlamovEPJB07}, where it was done in the
case of monolayer graphene.

The term $G_{\infty}(\omega)$, which is purely imaginary in the
considered spectral range, absorbs contributions from all
high-frequency core-level and valence band electronic
transitions, in particular, the ones involving $\sigma$ bands.
We assume that it does not contribute to the doping dependence
of the optical spectra.

The chemical potential $\mu$ is determined implicitly by the
doping level via equation:

\begin{equation}
\frac{1}{2\pi^2}\sum_{i}\int d^2\vec{q}
\left[f(\epsilon_{\vec{q},i}) - \frac{1}{2}\right]=n =
\alpha(V_{g} - V_{CN})\label{EqChemPot}
\end{equation}

\noindent We subtract $1/2$ in the brackets because the doping
level is counted relative to half filling of the $\pi$ bands.
The spin degeneracy is already included in equations
(\ref{EqGD}), (\ref{EqGIB}) and (\ref{EqChemPot}) but the
valley degeneracy is not. It appears that for the considered
range of energies and temperatures it is sufficient to perform
the momentum integration only in a circle of about 1 percent of
the total 2D Brillouin zone around the K point (and multiply by
2 to account for the valley degeneracy). In practical
implementation, we replace the integration with a summation
over $\sim 10^5$ $q$-points.

In equations (\ref{EqGD}) and (\ref{EqGIB}) we introduced two
different scattering rates for the Drude and the interband
components. Above 0.1 eV, the real part of the Drude
conductance is much smaller than the imaginary part, and the
latter is only very weakly affected by the Drude scattering in
this range. Since the data is not sensitive to the Drude
scattering rate, we adopted $\Gamma_{\mbox{\tiny D}}$ = 5 meV,
which corresponds to the value found in graphite
\cite{KuzmenkoPRL08}. We assume that the interband scattering
$\Gamma$ is constant, {\em i.e } it is energy, momentum and
band independent. By doing this, we neglected the energy
dependent electron-phonon and electron-electron scattering
processes, which is perhaps the most serious limitation of the
present model. We will see that it is likely in the origin of
some deviations of the model curves from the experimental ones.

Once the conductance of graphene is computed, the curves
$R_{gr}(\omega)$ can be calculated via the Fresnel equations
using the known optical constants of SiO$_{2}$ and Si, as
specified in the Appendix. The latter values are well known and
can be further refined by the fitting of the reflectivity of
the bare substrate $R_{sub}(\omega)$.

To summarize, the conductance $G(\omega, V_{g})$ and the
reflectivity $R_{gr}(\omega, V_{g})$ within the presented
approach depend on 9 parameters: $\gamma_{0}$, $\gamma_{1}$,
$\gamma_{3}$, $\gamma_{4}$, $\Delta$, $U$, $\Gamma$, $T$ and
$V_{CN}$. We applied the non-linear Levenberg-Marquardt
modelling routine\cite{NumericalRecipes} to directly fit the
experimental spectra. In order to speed up the iterations and
improve the convergence, the derivatives of the reflectivity
with respect to all adjustable parameters were calculated
explicitly using analytical formulas. One obstacle to this
approach is that the chemical potential should be determined
from the equation (\ref{EqChemPot}), which in general can be
done only numerically. Therefore we treated $\mu$ as a fitting
parameter and used equation (\ref{EqChemPot}) as a rigid
constraint of the least-square minimization.

\section{Results}\label{SectionResults}

\subsection{Reflectivity spectra and their tight-binding
modelling}\label{ResultsReflectivity}


\begin{figure*}[htp]
\includegraphics[width=6cm]{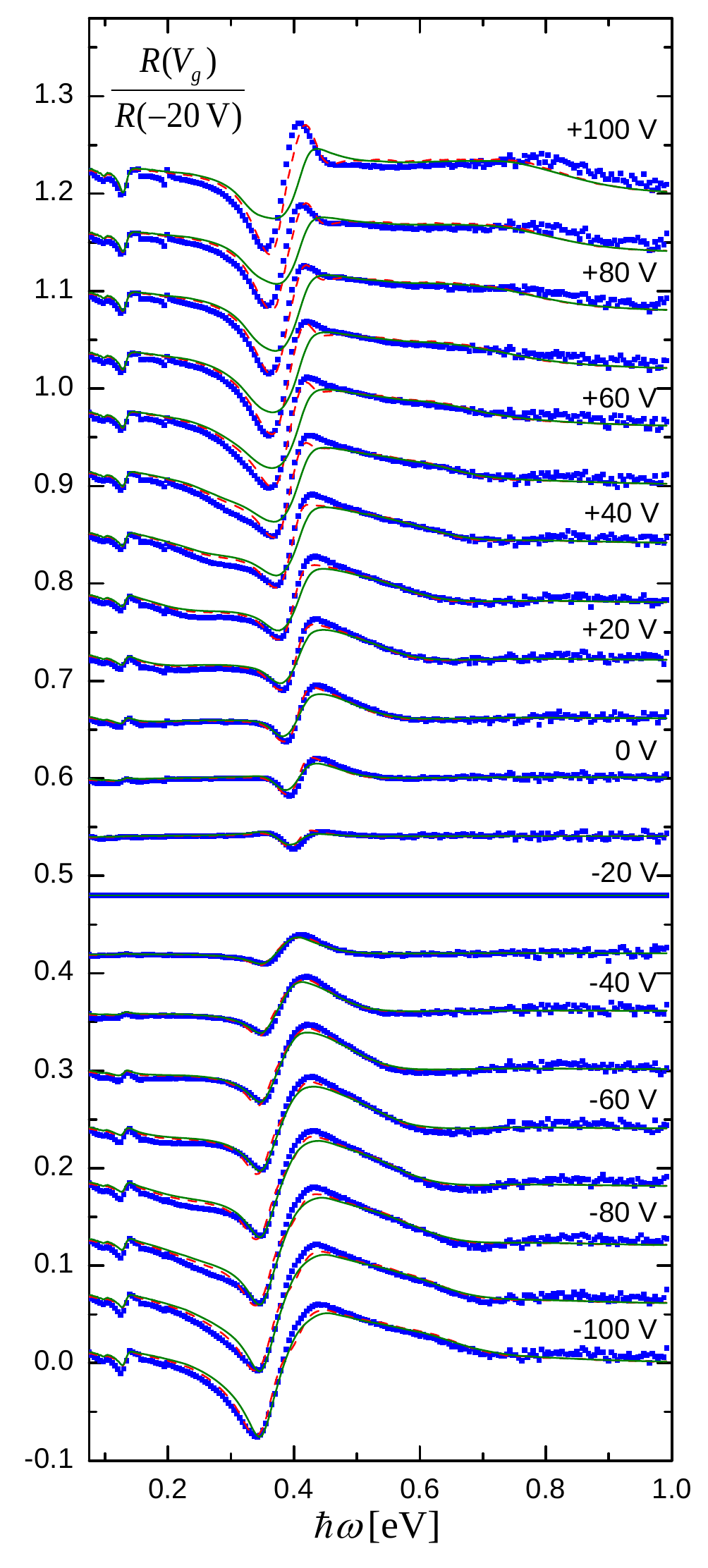}\includegraphics[width=6cm]{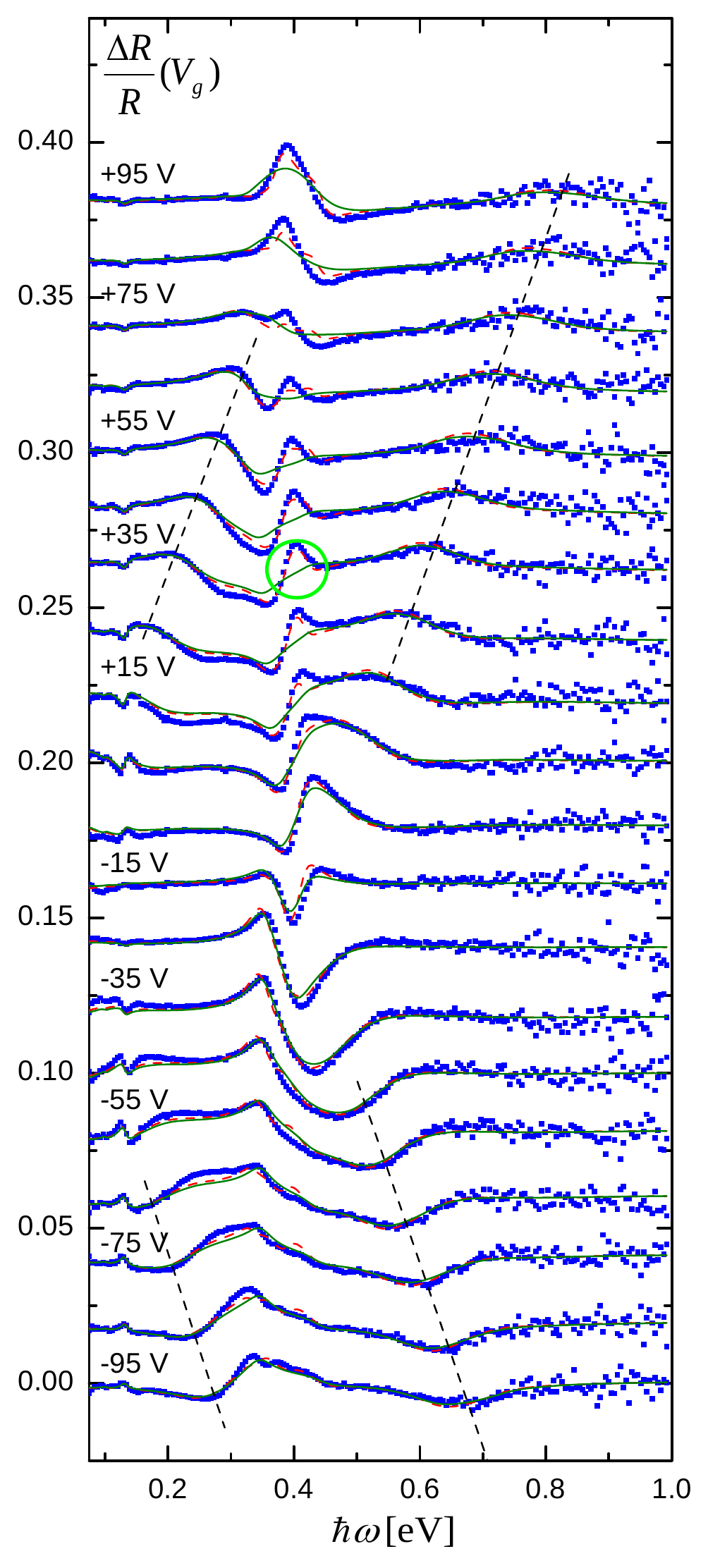}\includegraphics[width=6cm]{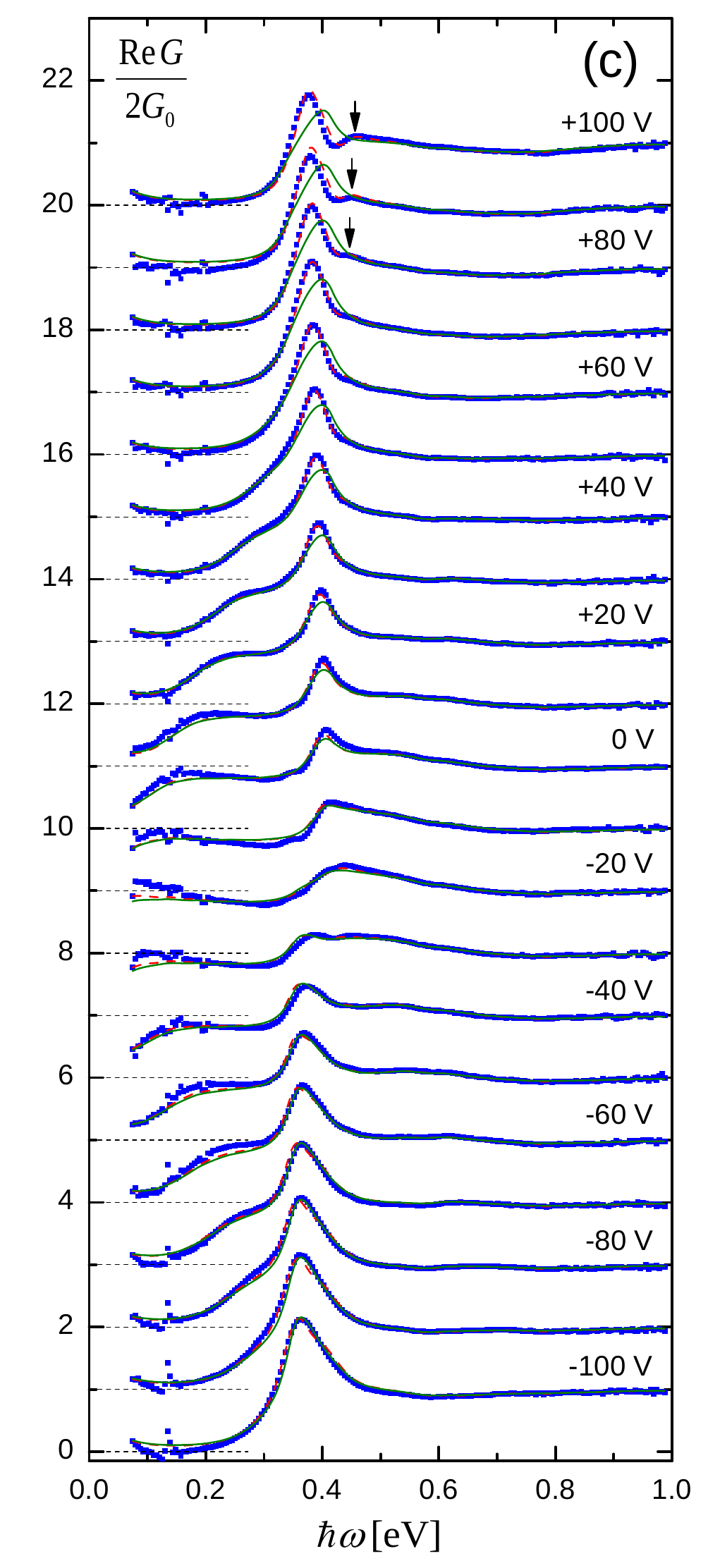}\\
\caption{(color on-line) (a) Self-normalized and (b) difference
reflectivity spectra as a function of $V_{g}$. Dots are the
experimental data points, solid and dashed lines represent
model fits 1 and 2 (without and with a bandgap) respectively.
(c) The real part of the optical conductance. Curves - the
result of the reflectivity fits (the colors and the line types
correspond to panels a and b), symbols - the refined
conductance obtained using a KK-consistent inversion as
described in the text. In all panels the curves are vertically
displaced for clarity.} \label{FigReflS1}
\end{figure*}

As it was discussed in the Section
\ref{SubsectionTechniquesSample}, the self-normalized
reflectivity $R(V_{g})/R(-20 V)$ (from now on we omit the index
"gr" for brevity) is the most accurately determined quantity,
which is therefore the best suitable for quantitative analysis.
These spectra, taken at the substrate temperature of 10 K, are
presented in Fig.\ref{FigReflS1}a for the whole span of gate
voltages used (from -100 V to +100 V with a step of 10 V). As
compared to Fig.\ref{FigSpectraMeas}c, the spectral resolution
in this figure is diminished to 5 meV. The spectra contain rich
structure that evolves in a peculiar fashion as a function of
the gate voltage. Such a complicated behavior is due to the
fact that all four bands are involved into the electronic
transitions that affect optical properties in the considered
energy range.

The amplitude of the structures of $R(V_{g})/R(-20 \mbox{ V})$
increases as the difference between $V_{g}$ and -20 V grows.
Therefore, it is useful to plot also the differential
reflectivity spectra (Fig.\ref{FigReflS1}b) defined as follows:
\begin{eqnarray}
\frac{\Delta R}{R}(\omega, V_{g})&\equiv& 2\frac{R(\omega,
V_{g}+5 \mbox{ V}) - R(\omega, V_{g}-5 \mbox{ V})}{R(\omega,
V_{g}+5 \mbox{ V}) + R(\omega, V_{g}-5 \mbox{ V})}.
\end{eqnarray}

\noindent Such a way of showing data emphasizes certain
structures, such as those indicated with dashed lines, and
their gate voltage dependence. Another advantage of this
representation is that it does not require {\em a priori} the
knowledge of the gate voltage corresponding to the
charge-neutral state.

\begin{table}
\caption{Parameter values obtained by the least-square fitting
of reflectivity spectra (fit 2). All parameters, except $T$ and
$V_{CN}$, are given in eV.}
\begin{tabular}{cccc}
  \hline
  \hline
  Parameter & This work & DFT calculation \cite{CharlierPRB92}\\
  \hline
  $\gamma_{0}$ & 3.16 $\pm$ 0.03& 2.598 $\pm$ 0.015\\
  $\gamma_{1}$ & 0.381 $\pm$ 0.003 & 0.34 $\pm$ 0.02\\
  $\gamma_{3}$ & 0.38 $\pm$ 0.06 &  0.32 $\pm$ 0.02\\
  $\gamma_{4}$ & 0.14 $\pm$ 0.03&  0.177 $\pm$ 0.025\\
  $\Delta$     & 0.022 $\pm$ 0.003 & 0.024 $\pm$ 0.01\\
  \hline
  $\Gamma$     & 0.018  $\pm$ 0.003 & - \\
  $T$          & 120 $\pm$ 15 K& - \\
  $V_{CN}$   & -22 $\pm$ 1 V & - \\

  \hline
  \hline
\end{tabular}
\label{TableFit}

\end{table}

We fitted the whole set of the $\Delta R/R$ spectra {\em
simultaneously} using the tight-binding parametrization,
described in Section \ref{SubsectionTechniquesModelling}. We
assumed that the $\gamma_{0}$, $\gamma_{1}$, $\gamma_{3}$,
$\gamma_{4}$, $\Delta$, $\Gamma$, $T$ and $V_{CN}$ do not
depend on $V_{g}$. Since one of our main goals is to detect and
measure the bandgap, we performed fits in two different ways.
First, we set the parameter $U$ to zero at all gate voltages
(fit 1) so that the difference between the spectra is only due
to a variation of the chemical potential. In the second run
(fit 2), the bandgap was allowed to vary as a function of
$V_{g}$ in such a way that $U$ at each value of the gate
voltage was treated as an independent parameter. In each case,
we tried different sets of initial parameters (within the scope
of physically reasonable values) and checked that the fitting
routine converges to the same result. The parameter confidence
limits were estimated based on the correlation analysis
\cite{NumericalRecipes} and by repeating the process after
varying data points within their error bars.

The model curves corresponding to the fits 1 and 2 are shown in
Figures \ref{FigReflS1}a and \ref{FigReflS1}b (solid green and
dashed red lines respectively). Both fits show almost the same
match to the data outside the region around 0.4 eV. However,
within this region the fit 1, which does not involve the
bandgap, is qualitatively worse. It fails to reproduce some
strong structures, in particular, the ones marked with the
circle. As discussed in
Refs.\onlinecite{NicolCarbottePRB08,ZhangPRB08,KuzmenkoPRB09},
this is exactly the region, where the bandgap is expected to
affect the spectra. At the same time, the quality of the fit 2
is remarkably good. Thus our data unequivocally show the
presence of the bandgap. There are still some mismatches that
we shall address separately.

The parameters of the fit 2, apart from $U$, which depends on
the gate voltage, are given in the Table \ref{TableFit}. One
can see that the SWMcC parameters can be determined from the
infrared spectra. Except $\gamma_{3}$, these parameters were
already determined in previous infrared
studies\cite{WangScience08,ZhangPRB08,LiPRL09,KuzmenkoPRB09},
by monitoring the gate voltage dependence of easily
recognizable spectral features, such as the maximum of the
$\gamma_{1}$ peak. Using the least-square fitting method, the
parameter $\gamma_{3}$ can now also be estimated. This term
results in a deformation of the $\gamma_{1}$ peak, but its
effect on the spectra is more complicated than just a
broadening produced by $\Gamma$ (that we find to be about 18
meV). One should mention, that in the fit 1 (where the bandgap
was not included) this parameter relaxed to an artificially
large value (about 0.5 eV), in order to mimic somehow the
bandgap induced smearing of the $\gamma_{1}-peak$.

For comparison, we also reproduce the SWMcC parameters obtained
from a density functional theory (DFT) calculation on graphite
by Charlier et al. \cite{CharlierPRB92} (a more complete
overview is given, for example in Ref.\onlinecite{ZhangPRB08}).
The agreement is good, except for $\gamma_{0}$, for which the
DFT values are somewhat lower. Nevertheless, $\gamma_{0}$
deduced from various experiments on graphite
\cite{DresselhausAdvPhys81} is in an excellent agreement with
our result. Based on the obtained value of $\gamma_{0}$, we
find the Fermi velocity $v_{F} = (3/2)a\gamma_{0}/\hbar$ to be
$1.02 \pm 0.01 \times 10^6$ m/s ($a$ = 1.42 \AA\ is the
nearest-neighbor interatomic distance).

The gate voltage corresponding to zero doping, $V_{CN}$, can be
determined with a very good accuracy. As it was mentioned in
Section \ref{SubsectionTechniquesSample}, this value is
slightly different from the one extracted from the simultaneous
transport measurement. It is possible that in the case of large
flakes and a non-optimal geometry of electrical contacts, it
gives a more accurate value than the one given by the maximum
of the DC resistivity as it is not affected by the distribution
of the measurement currents in the flake. Infrared spectroscopy
can be used therefore as an independent indicator of the doping
level.

We find the broadening $\Gamma$ to be about 15-20 meV. This
value is larger than the one reported in
Ref.\onlinecite{ZhangPRB08} ($\approx 0.02\gamma_{1} =$ 8 meV),
which is perhaps due to a higher concentration of charging
impurities (the ones that shift the charge neutral point from
the zero bias). At the same time is considerably smaller than
the broadening of about 60 meV found in
Ref.\onlinecite{ZhangNature09} on double gated graphene, which
is probably related to extra scattering and/or inhomogeneity
introduced by the top gate.

At first surprisingly, the deduced temperature of graphene $T$
is of the order of 100 K, even though the substrate was kept at
10 K. Although the true graphene temperature may indeed be
somewhat higher due to a weak thermal contact between the
warped flake and the substrate, another plausible explanation
is that this is an indication of the spatial inhomogeneity of
the chemical potential. It is easy to see that if we neglect
the change of the bands as a function of $\mu$ then a smearing
of the chemical potential $\mu\rightarrow\mu \pm\delta\mu$ has
almost the same effect in the Kubo formula as increasing the
temperature ($k_{B}T_{eff}\sim\delta\mu$). Thus we get an upper
limit of about 10 meV to the inhomogeneity of the chemical
potential. It is worth emphasizing that the thermal and the
scattering induced broadening given by $T$ and $\Gamma$
respectively are clearly distinguishable by the fitting
routine.


\begin{figure}[htp]
\includegraphics[width=7cm]{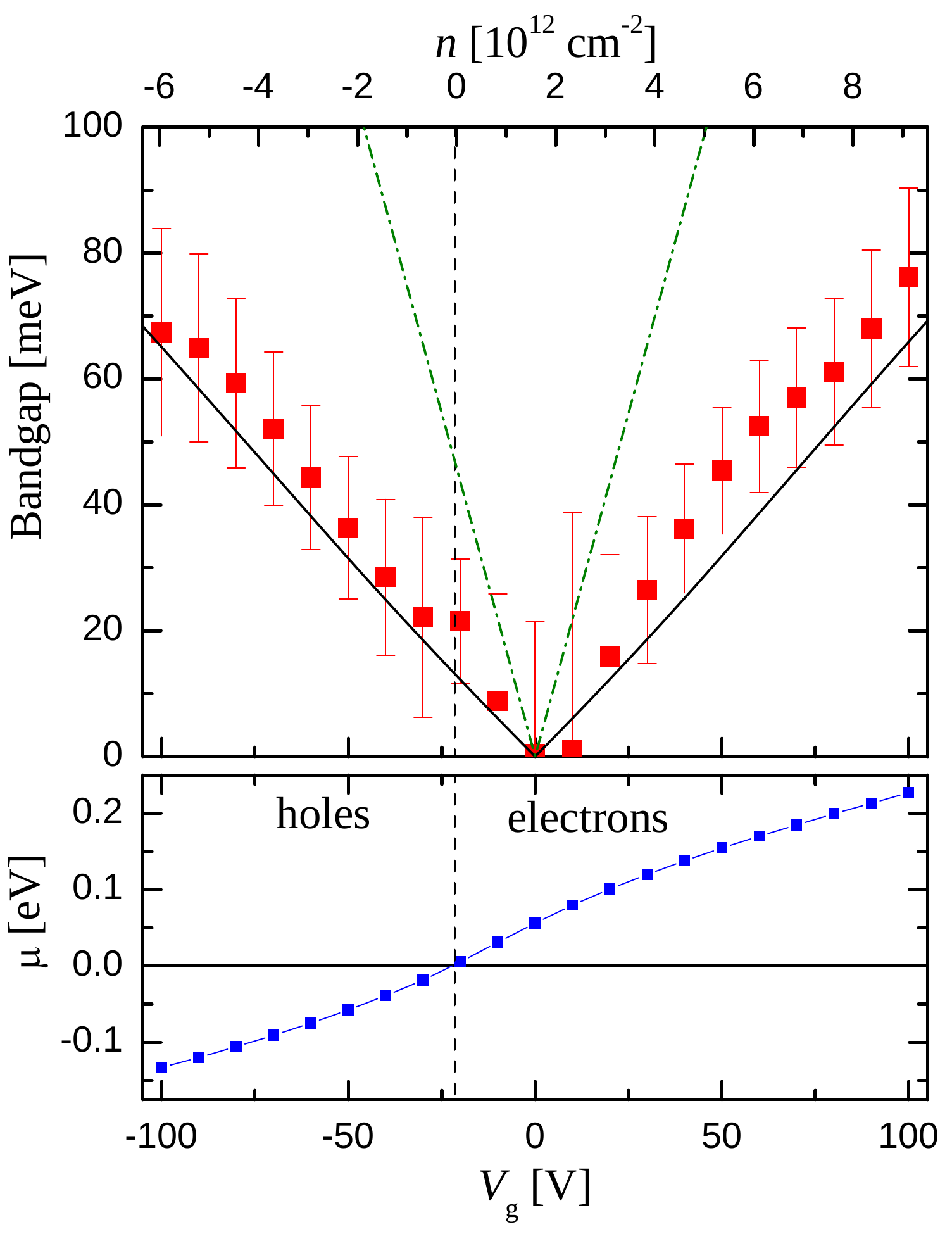}\\
\caption{(color on-line) The extracted values of $|U|$ (top
panel) and the chemical potential (bottom panel) as functions
of the gate voltage and doping. Dash-dotted line - the
calculated "unscreened" value of $|U|$, solid line - an {\em ab
initio} DFT calculation from Ref.\onlinecite{GavaPRB09}, which
takes screening effects into account.} \label{FigGap}
\end{figure}

Figure \ref{FigGap}a shows the extracted bandgap as a function
of the gate voltage and doping. Here we take $|U|$ as a measure
of the bandgap, since it practically coincides with
$\Delta_{g}$ in the considered doping range. One can see that
at small gate voltages the bandgap goes to zero within error
bars and it grows almost linearly for both electron and hole
doping, reaching 70-80 meV at the maximum applied gate
voltages. Interestingly, the minimum is closer to zero gate
voltage (where $n \approx 2\times 10^{12}$ cm$^{-2}$) than to
$V_{g} = V_{CN}$ ($n = 0$). As discussed by Castro et
al.\cite{CastroPRL07}, such a shift can be understood by
considering the dopant molecules adsorbed by the surface acting
as an effective top gate electrode. However, in this case, the
zero-gap point is expected to be at $V_{g} = -V_{CN}$. Seeing
zero gap at zero gate voltage would be expected if the dopants
are intercalated between the carbon layers, so that they do not
introduce an interlayer electrostatic asymmetry, however we do
not have any independent experimental verification of this
happening. If the dopants were below the flake, in this picture
one expects the gap to vanish at the charge neutral point.


\begin{figure}[htp]
\includegraphics[width=8.5cm]{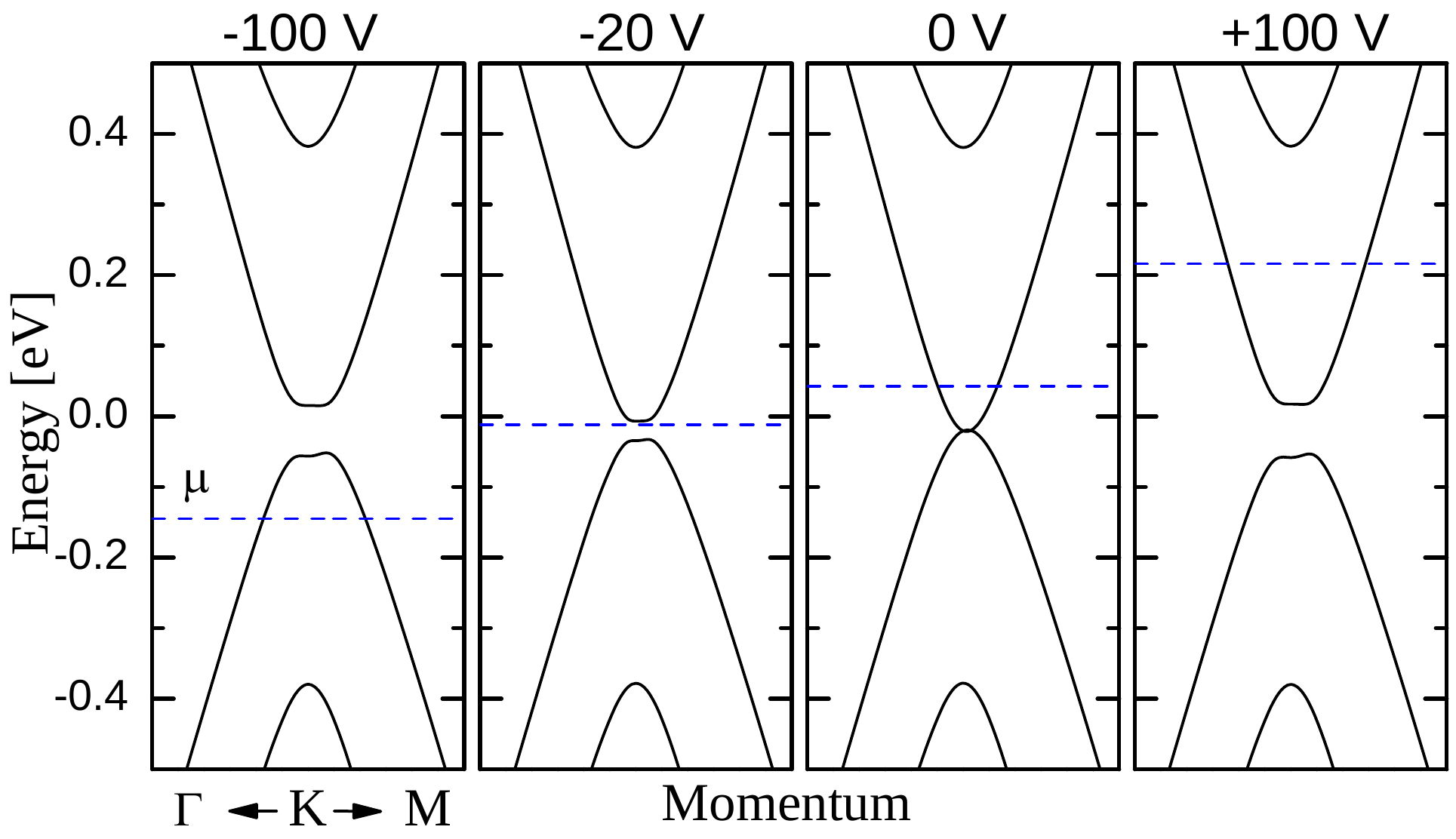}\\
\caption{(color on-line) Electronic bands (calculated using
parameters extracted from optical spectra) and chemical
potential at selected gate voltages.} \label{FigBandPicture}
\end{figure}

From Figure \ref{FigGap}b one can see that the chemical
potential shows a monotonic, slightly sublinear increase with
doping. In figure \ref{FigBandPicture} the calculated band
structures, corresponding to the gate voltages -100, -20, 0 and
+100 V, are presented, together with the position of the
chemical potential.

\subsection{Optical conductance}

Fig.\ref{FigReflS1}c shows the real part of the optical
conductance $G(\omega)$ calculated using the obtained
tight-binding parameters as described above. As in
Fig.\ref{FigReflS1}a, green solid and red dashed lines
correspond to the fits 1 and 2 respectively. The conductance is
normalized by $2G_{0}$, which is the theoretical asymptotic
value for bilayer graphene at high frequencies
\cite{AbergelFalkoPRB07,NicolCarbottePRB08}. One can see that
the model curve corresponding to the fit 1 shows a very broad
$\gamma_{1}$ at positive gate voltages because the parameter
$\gamma_{3}$ in this fit was unrealistically large, as
discussed above.

Since some deviations between the model fits and the experiment
are present, we refined optical conductance using the
Kramers-Kronig constrained variational method
\cite{KuzmenkoRSI05}. Within this approach, we represented the
conductance as a sum of the two terms: $G(\omega) =
G_{mod}(\omega) + G_{var}(\omega)$. The model term is
calculated using equations (\ref{EqG}), (\ref{EqGD}) and
(\ref{EqGIB}). $G_{var}(\omega)$ is a variational
Kramers-Kronig constrained correction, needed to reproduce all
remaining fine details of the experimental reflectivity
spectra. \cite{KuzmenkoRSI05}. At the refinement stage,
$G_{mod}(\omega)$ was fixed and $G_{var}(\omega)$ was adjusted
in order to get the perfect match to the reflectivity spectra.
The refined conductivity is shown in Fig. \ref{FigReflS1}c with
symbols. Since we based our analysis on the relative
reflectivity spectra, this procedure gives most accurately the
{\em relative} changes of $G(\omega)$ as a function of $V_{g}$
and $\omega$ (the accuracy is better than $0.1 G_{0}$), while
the error bars for the absolute level of $G(\omega)$ can be
somewhat larger. This explains slightly negative values of
$\mbox{Re }G(\omega)$ at low frequencies at high gate voltages.
Weak structures below 0.15 eV are artefacts coming from the
optical phonons in SiO$_{2}$, which are not fully cancelled in
the fitting procedure, probably due to a weak dependence of
these phonons on the electric field, which is not included in
our model.

\subsection{Temperature dependence}

\begin{figure}[htp]
\includegraphics[width=8.5cm]{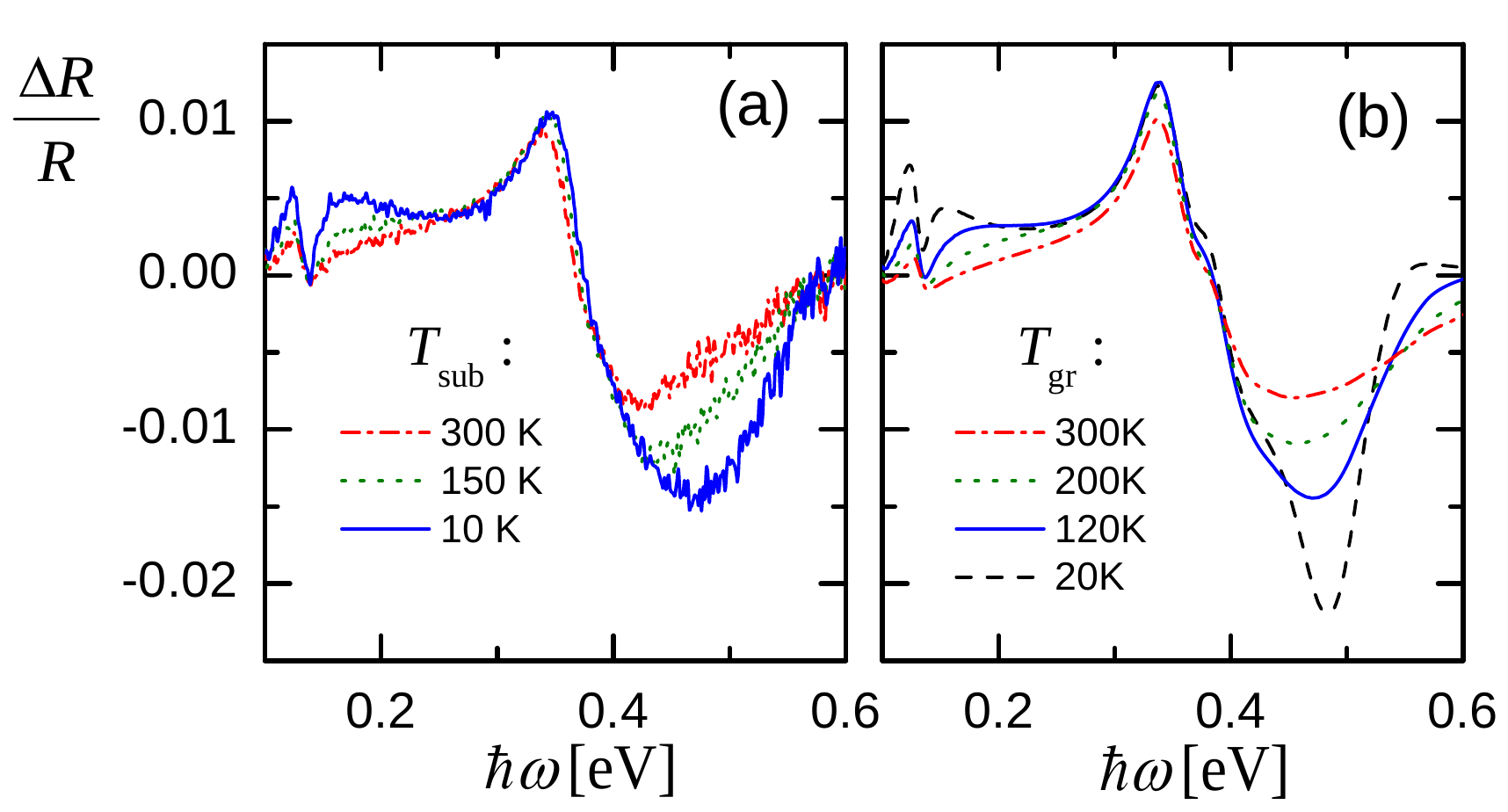}\\
\caption{(color on-line) (a) The difference reflectivity at
$V_{g} = -45$ V at three temperatures of the substrate: 300 K,
150 K and 10 K. (b) Model curves of the same quantity,
calculated by varying the effective temperature of graphene and
keeping other parameters unchanged (shown in Table
\ref{TableFit}).} \label{FigTDep}
\end{figure}

In addition to tuning the spectra by the gate voltage, varying
the temperature provides another important piece of
information. As an example, figure \ref{FigTDep}a shows $\Delta
R(\omega, -45 V)/R$ taken at the {\em substrate} temperature
$T_{sub}$ of 10 K, 150 K and 300 K. The spectra clearly change
with cooling down. In general, the structures are getting
sharper at low temperatures. However, the sharpening is far
from being simply a uniform broadening, such as due the
electronic scattering (parameter $\Gamma$). In particular, the
peak at 0.35 eV does not change, while the dip at 0.45 eV and a
structure at 0.2 both show a pronounced temperature variation.


Panel (b) shows calculations of the same quantity at the
following set of {\em graphene} temperatures: 20 K, 120 K 200 K
and 300 K. In all cases, the same parameters, except $T$, were
used (Table \ref{TableFit}). The calculated temperature
dependence reproduces very well the experimental one if one
assumes that the effective temperature of graphene is higher
than $T_{sub}$. As we discussed above, this temperature
mismatch may be in part due to the spatial broadening of the
chemical potential. One can see that at $T$ = 20 K (dashed
line), the spectral structures are expected to be much sharper
than at 120 K. Similar spectra comparisons at other gate
voltages (not shown) provide the same results.

The observation that the effect of temperature on the spectra
is highly frequency selective is explained by the fact that
only electronic transitions, for which either initial or the
final state are close to the Fermi level, are affected by the
temperature. As one can anticipate from the good match between
spectra in panels (a) and (b) of Fig.\ref{FigTDep}, the
application of the same least-square fitting procedure at
higher temperatures provides model parameters, which are
essentially the same as the ones presented. Therefore we focus
largely on the low temperature results. Nevertheless, the
ability to quantitatively predict spectra at high temperatures
based on the results obtained at low temperature corroborates
the consistency of the used model approach.

\section{Discussion}\label{SectionDiscussion}

\subsection{Seeing bandgap optically: zero versus finite doping}



\begin{figure}[htp]
\includegraphics[width=8.5cm]{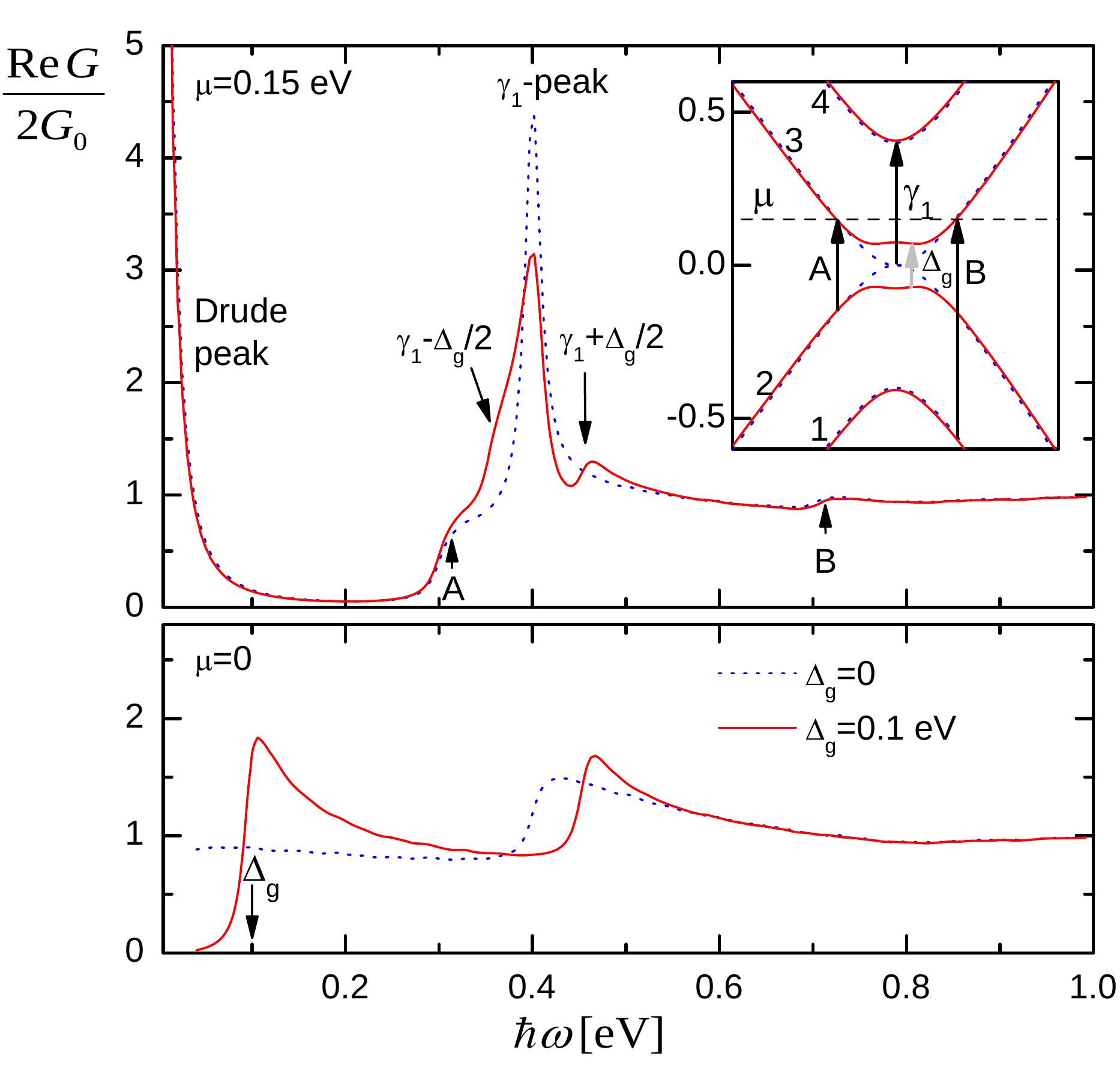}\\
\caption{(color on-line) Model demonstration of the effect of
the bandgap on optical spectra in the case of finite doping and
the case of zero doping. Top panel: the chemical potential is
larger than the bandgap; bottom panel: $\mu$ is zero. The inset
shows the corresponding bands (energy, in eV, versus momentum
near the K point). A simplified set of SWMcC parameters is
used: $\gamma_0$ = 3 eV, $\gamma_1$ = 0.4 eV, $\gamma_{3} =
\gamma_{4} = \Delta = 0$. The broadening $\Gamma$ is 0.01 eV.}
\label{FigExplanation}
\end{figure}

Optical spectroscopy is routinely used for the bandgap
measurements in usual semiconductors due to the fact that only
the photons, which energy exceeds the bandgap, are absorbed by
electron-hole excitations. However, bilayer graphene is special
in a sense that the bandgap is not intrinsic but is induced by
the gate voltage. Therefore, if only one gate is attached to
the flake, as in the present case, then the gap opening is
inevitably accompanied by doping. Making two gate electrodes
(on top and on bottom) would induce the bandgap without doping
\cite{OostingaNatMat08}. Figure \ref{FigExplanation}
demonstrates, using a simplified set of SWMcC parameters, the
effect of the bandgap in both cases. The bottom panel
corresponds to the undoped case ($\mu = 0 $). One can see that
an infrared absorption threshold appears at $\hbar\omega
\approx \Delta_{g}$, due to the transitions between the bands 2
and 3 across the bandgap (the bands are numbered in the inset).
Such a structure was observed indeed in a recent paper by Zhang
et al.\cite{ZhangNature09}, who used a double-gate bilayer
graphene device. The second notable effect is a shift by
$\Delta_{g}/2$ of the second threshold at $\gamma_{1}$
corresponding to the transitions between bands 1 and 3 and also
to ones between bands 2 and 4.

The top panel describes the case of a finite chemical potential
(electron doping). There is a striking difference between two
cases. Now the opening of the gap does not produce an
absorption threshold, since the transitions across the gap are
blocked by the Pauli principle. However, the gap affects the
lineshape of the $\gamma_{1}$ peak that originates from a
combination of interband transitions $3 \rightarrow 4$ and $2
\rightarrow 4$. Most notably, a satellite peak at about
$\gamma_{1}+\Delta_{g}/2$ shows up. In addition, a shoulder at
about $\gamma_{1}-\Delta_{g}/2$ appears. The satellite and the
shoulder stem from the transitions $2 \rightarrow 4$ and $3
\rightarrow 4$ respectively close to the K point, where the are
separated by the bandgap energy. Thus, in the doped case the
only way to measure the bandgap is to analyze the shape of the
$\gamma_{1}$ peak.

In reality, the position and the shape of the peak are also
affected by the parameters $\gamma_{3}$, $\gamma_{4}$ and
$\Delta$, not included in the above demonstration, and further
broadened by electronic scattering. Therefore, when the gap is
small, its extraction from the optical spectra requires direct
fitting of the data using a complete set of SWMcC parameters.
When the gap is large, the identification of the gap becomes
easier as the satellite to the main peak is more pronounced. We
note that in actual data one can clearly recognize a satellite
to the $\gamma_{1}$ peak for $V_{g} \geq$ 80 V (shown by arrows
in Fig.\ref{FigReflS1}c).

Although the fit reveals the presence of the bandgap also at
negative gate voltages (Fig.\ref{FigGap}a), the conductance
spectra do not show a clear satellite at this doping side. Such
a difference is in part due to the electron-hole asymmetry,
which results in a stronger broadening of the $\gamma_{1}$ peak
at the hole doping, and in part due to the shift of the charge
neutrality point. However, in section
\ref{SubsectionDiscussionMapping} we shall demonstrate direct
signatures of the bandgap for both polarities of the gate
voltage.

\subsection{Bandgap: the role of self-screening}

As discussed in section \ref{SubsectionTechniquesModelling},
the bandgap is determined by the parameter $U$, which is
defined as the difference of the electrostatic potential on the
two layers. As it was extensively discussed in the literature
\cite{McCannPRB06,CastroPRL07,AokiAmawashiSSC07,MinPRB07,ZhangPRB08,FalkovskyCM09},
the self-screening of the external field plays a crucial role
in the determination of $U$. Our data fully agree with this.
One can see (Figure \ref{FigGap}) that the experimental value
of the bandgap is more than two times smaller than the
"unscreened value" (dashed-dotted line), given by the external
field multiplied by the interlayer distance (3.35 \AA)
\cite{GavaPRB09} (we assumed that the bandgap vanishes at
$V_{g} = 0$, based on the experimental results, which means
that charging impurities do not introduce any imbalance of the
interlayer potential). The same observation was made in
Ref.\onlinecite{ZhangNature09}

A proper microscopic calculation of the bandgap must be done
self-consistently, since the screening depends on the bandgap
and vice versa. Such a complicated problem was treated on the
Hartree level based on the tight-binding model
\cite{McCannPRB06,CastroPRL07,ZhangPRB08,FalkovskyCM09} as well
as using {\em ab-initio} methods
\cite{AokiAmawashiSSC07,MinPRB07,GavaPRB09}. These calculations
provide the doping dependent bandgap, which is much
closer\cite{McCannPrivateCommunication} to the present
experiment than the "unscreened" model. As an example, we
present on Figure \ref{FigGap} the {\em ab-initio} DFT
calculation of Gava et al. \cite{GavaPRB09} (solid line), which
shows a good agreement with the experimental data.

\subsection{"Photon energy - gate voltage" mapping of the interband transitions}
\label{SubsectionDiscussionMapping}

\begin{figure*}[htp]
\includegraphics[width=18cm]{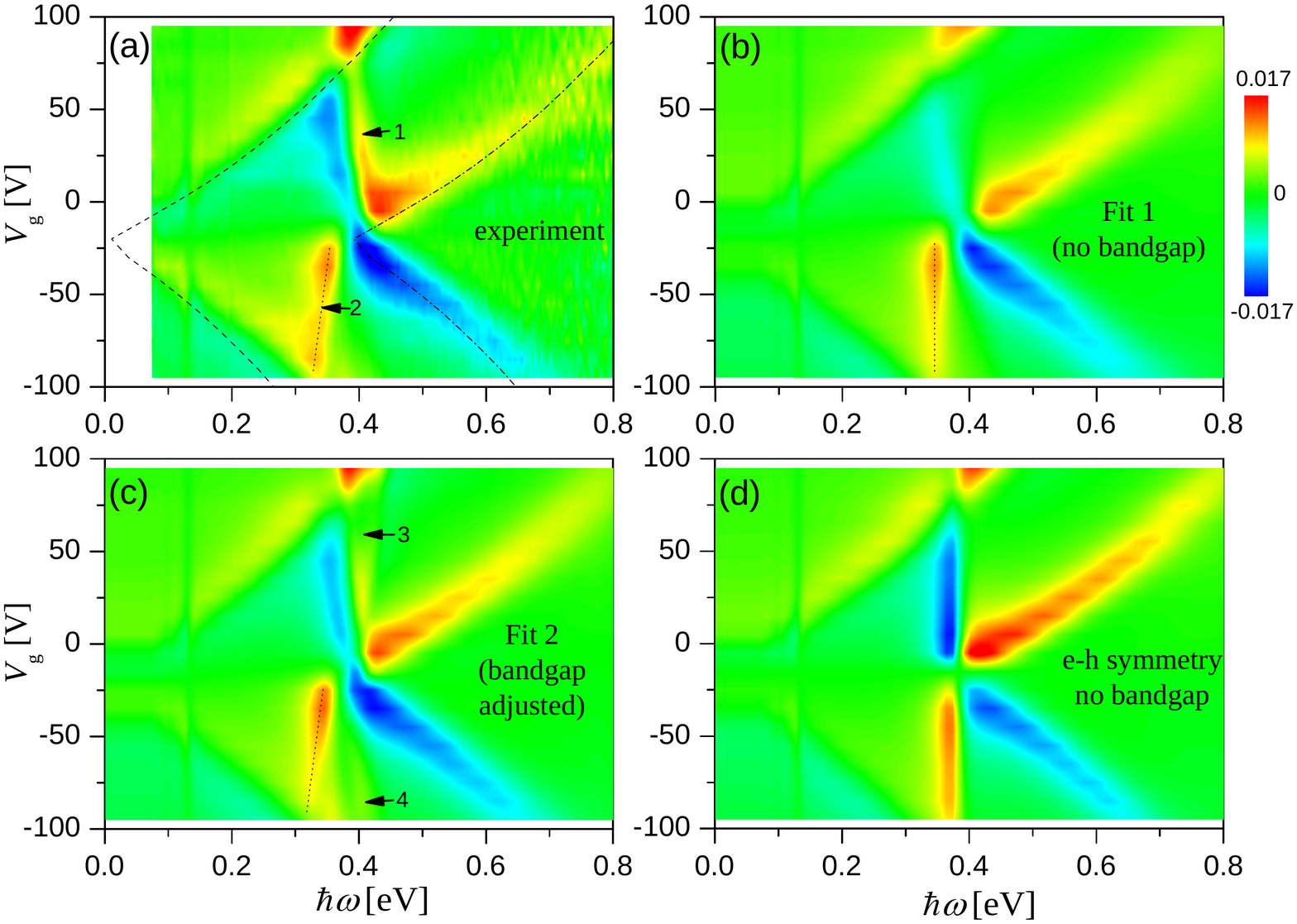}\\
\caption{(color on-line) Color maps of $(\Delta
R/R)(\omega,V_{g})$. (a) - experiment. (b) - model curves
corresponding to Fit 1 (zero bandgap), (c) - model curves
corresponding to Fit 2 (including bandgap), (d) - model curves
in the absence of the bandgap and electron-hole asymmetry. The
color scheme is the same in all graphs. The dashed and
dash-dotted lines in panel (a) are the dependencies of
$\omega_{A} = 2|\mu|$ and $\omega_{B} = \gamma_{1}+ 2|\mu|$
respectively.} \label{FigColorPlotDRR}
\end{figure*}

We saw that the extraction of the optical conductance from the
measured spectra is rather involved in the present case, where
the measured spectra depend on both real and imaginary parts of
$G(\omega)$ (as detailed in the Appendix). Now we propose a
simple way to visualize electronic transitions based on the
{\em raw} reflectivity data, which most clearly demonstrates
the electron-hole asymmetry, the opening of the bandgap and
other feature of the band structure.

In figure \ref{FigColorPlotDRR}a, the whole set of experimental
spectra $\Delta R/R$ is represented as a color map in the
coordinates $(\hbar\omega;V_{g})$. One can see a set of lines
that resemble somewhat band dispersions seen in ARPES. First we
note the two "$<$"-like structures, shifted with respect to
each other along the photon energy axis. They correspond to the
onset-like features in the optical conductance, marked in
Fig.\ref{FigExplanation} as A and B and related to the
interband transitions $2 \rightarrow 3 $ and $1 \rightarrow 3$
respectively ($2 \rightarrow 3 $ and $2 \rightarrow 4$ in the
case of hole doping). Indeed, one can see that they match
closely the  expected threshold energies $\omega_{A}(V_{g}) =
2|\mu(V_{g})|$ (dashed line) and $\omega_{B}(V_{g}) =
\gamma_{1} + 2|\mu(V_{g})|$ (dotted line). Although several
experimental papers presented infrared spectra of gated bilayer
graphene
\cite{HenriksenPRL08,WangScience08,LiPRL09,MakCM09,KuzmenkoPRB09,ZhangNature09},
the second threshold was reported only in
Ref.\onlinecite{KuzmenkoPRB09}. Here we reaffirm, based on a
new set of data, the existence of the second threshold, which
is essential for the overall consistency of the tight-binding
approach.

In figure \ref{FigColorPlotDRR}a the presence of electron-hole
asymmetry is quite obvious, since in the case of perfectly
symmetric bands with respect to the Dirac point the $\Delta
R/R$ spectra should be precisely {\em antisymmetric} with
respect to $V_{g} = V_{CN}$, as exemplified in the hypothetical
graph of Fig. \ref{FigColorPlotDRR}d. Within the SWMcC model,
the asymmetry between electron and holes bands is due to the
hopping term $\gamma_{4}$ and the on-site energy difference
$\Delta$. In Refs. \onlinecite{ZhangPRB08,LiPRL09} they were
deduced from the doping dependence of the position of the
maximum of the $\gamma_{1}$-peak. Since the maximum location is
affected not only by $\gamma_{1}$, $\gamma_{4}$ and $\Delta$
but also by $\gamma_{3}$, $\Gamma$ and, most importantly, by
$U$, we choose to determine all parameters, including
$\gamma_{4}$ and $\Delta$, by fitting of the whole set of
spectra.

This way of presenting spectra also allows us to see distinct
features related to the opening of the bandgap. These features
appear to be quite different on the two doping sides due to the
electron-hole asymmetry. In panels (b) and (c) of the same
figure, we show the fits of $\Delta R/R$ without and with the
bandgap respectively (namely, fits 1 and 2 described in Section
\ref{ResultsReflectivity}). On the electron side ($V_{g}
> V_{CN}$) the 'ridge' indicated as '1' finds absolutely
no counterpart in the fit 1, but is mimicked by in the fit 2.
On the hole side, the ridge marked as '2' clearly disperses
towards low frequencies as the absolute value of $V_{g}$ is
increasing. This trend is well captured by the fit 2, while in
the fit 1 this ridge is precisely vertical.

\subsection{Deficiencies of the tight binding description with constant scattering}

\begin{figure}[htp]
\includegraphics[width=8.5cm]{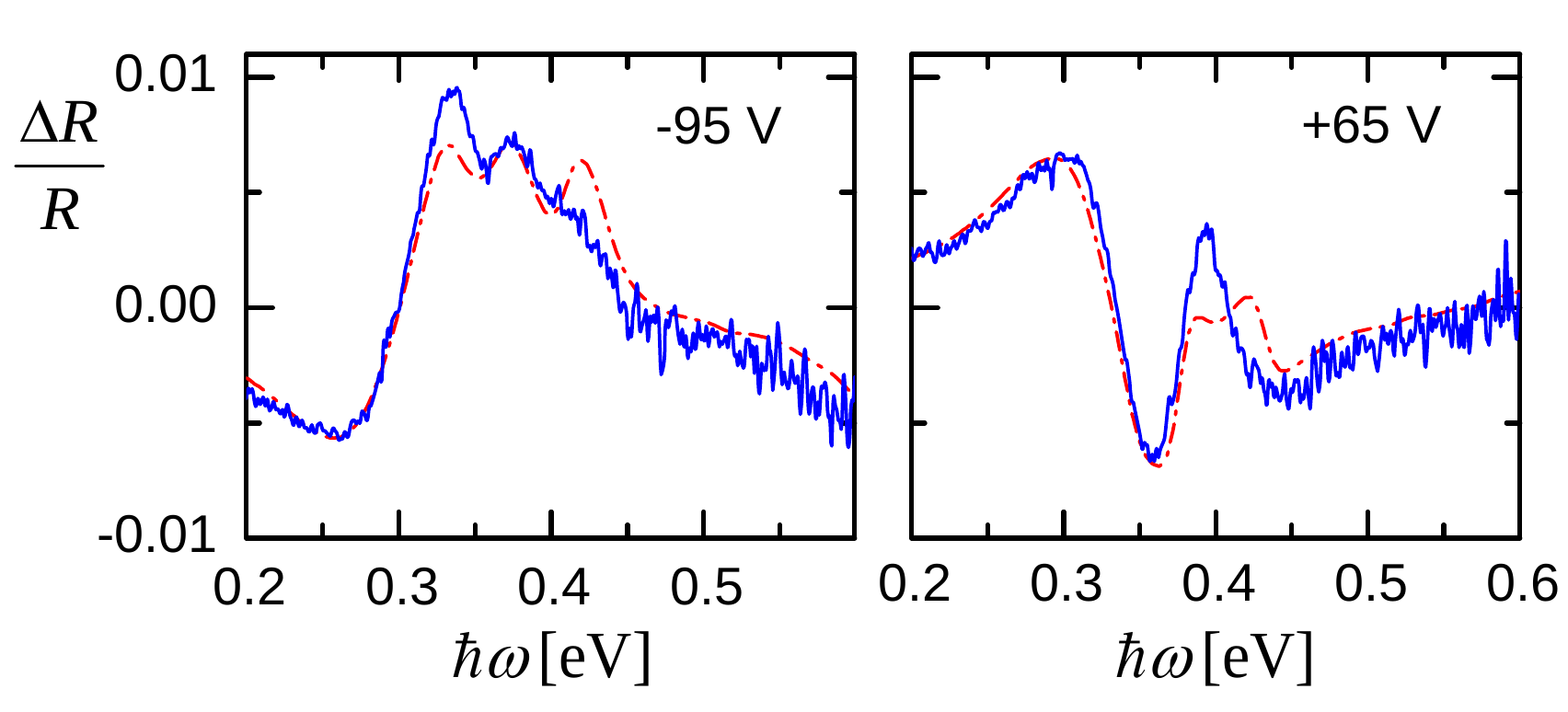}\\
\caption{(color on-line) The experimental curves $(\Delta
R/R)(\omega)$ (solid lines) at $V_{g}$ = -95 V (left panel) and
+65 V (right panel), together with the best fitting curves
(dash-dotted lines).} \label{FigFitDetails}
\end{figure}

Although the overall agreement between the panels (a) and (c)
of Fig. \ref{FigColorPlotDRR} is very good, a closer inspection
reveals some deficiencies of the fit 2. For the electron
doping, at gate voltages between 50 and 80 V, the ridge
indicated by 1 is quite narrow in the experiment but is broad
and barely recognizable in the fit 2 (as indicated by 3 in the
panel (c). For the hole doping, the fit 2 contains a weak extra
ridge (marked as '4') which is not clearly present in the
experiment. In Fig.\ref{FigFitDetails} we concentrate on these
doping levels, taking the gate voltages $V_{g}$ = -95 V and +65
V as examples. Here we improved the match even further as
compared to Fig.\ref{FigReflS1}b by slightly compromising the
fit quality at other gate voltages (although all the bandgaps
were kept the same).


The model curve at -95 V t is featured by three distinct peaks
between 0.3 and 0.45 eV. Crudely speaking, they are related
respectively to the shoulder, the main maximum and the
satellite to the $\gamma_{1}$ peak depicted in
Fig.\ref{FigExplanation}. The value of $U$ is related to but
somewhat smaller than the distance between the leftmost and the
rightmost peaks. We can see that the data also show three peaks
at the same photon energies, but the first peak is stronger and
the third one is much weaker than the counterpart on the
theoretical curve (the latter one gives rise to the feature '4'
in Fig.\ref{FigFitDetails}c).

At $V_{g} = $+65 V the peak at 0.4 eV is rather sharp in the
experiment but has a pronounced double structure in the model.
Although this does not question the existence of the bandgap as
such (recall that the best fit without the bandgap does not
show this peak at all), such a discrepancy is too significant
to be ignored.

Presently, the origin of the shown mismatches is not clear.
This may be an indication that more hopping terms need to be
taken into account. In this respect, it would be instructive to
compare optical spectra directly to the results of {\em
ab-initio} band structure calculations. Another possibility is
that the discrepancies are caused by our assumption that the
scattering $\Gamma$ is the same for all electronic states,
which can fail due to the electron-phonon and electron-electron
interactions. Last but not least we have assumed in our
analysis a rigid band model, i.e. the tight binding band
parameters (apart from the bandgap) are assumed to be
independent of doping and gate voltages. That this may not be
strictly the case was experimentally shown in an ARPES study on
epitaxial graphene\cite{OhtaScience06}, where an increase of
$\gamma_{1}$ by about 3 \% was observed, when $U$ changed from
0 to 100 meV. Electron-correlation effects introduce a doping
and energy dependent renormalization of the bare dispersion.
Also the gate voltages influence the interatomic tunneling
matrix elements, which in turn affect the tight binding
parameters. Studying the manifestation of these interactions in
optical spectra will undoubtedly be one of the most intriguing
directions in the further research of graphene.

\section{Summary}

We presented a detailed analysis of infrared reflectivity
spectra of bottom gated bilayer graphene that allowed us to
determine the tight-binding Slonczewski-Weiss-McClure
parameters and the doping dependence of the bandgap induced by
the electric field generated by the gate. The direct
least-square fitting of the whole set of infrared spectra using
the SWMcC Hamiltonian and the Kubo formula turns out to be a
very efficient technique to disentangle the complicated
interplay of various band structure parameters in the optical
spectra. It also provides independent information about the
extrinsic doping level.

Our analysis clearly shows the presence of the bandgap, which
depends almost linearly on the gate voltage. This dependence
agrees with the tight-binding and {\em ab initio} calculations
that take the screening of the external field by the $\pi$
bands into account. At the maximum applied gate voltage of 100
V the bandgap reaches about 80 meV, which is three times larger
than $k_{B} T$ at room temperature. Even higher values of the
bandgap (up to 250 meV) could be obtained on double-gated
bilayer graphene \cite{ZhangNature09}, making this material
very promising for applications.

The very fact of achieving quantitatively good fits is a strong
indication that the tight-binding model is quite accurate for
the actual band structure of bilayer graphene. Nevertheless,
some discrepancies remain, and further investigations will be
needed to explore their origin in the context of
electron-phonon and electron-electron interactions.

We are grateful to A. K. Geim, A. H. MacDonald, F. Mauri, E. McCann, L.A. Falkovsky  and
D.N. Basov for illuminating discussions. This work was
supported by the Swiss National Science Foundation (SNSF) by
the grant 200021-120347, through the National Center of
Competence in Research "Materials with Novel Electronic
Properties-MaNEP".

\section*{APPENDIX: Relation between the reflectivity and the optical conductance of graphene}

The reflectivity of bare substrate, and graphene on top of the
substrate (Fig.\ref{FigSpectraMeas}a) can be calculated based
on the optical conductance of graphene $G(\omega)$ and the
known dielectric functions $\epsilon(\omega)$ of SiO$_{2}$ and
Si \cite{BlakeAPL07}. We can treat the silicon layer as
semi-infinite, since in our case it is thicker than the
penetration depth. The Fresnel equations for the reflectivities
can be written as follows:

\begin{eqnarray}
R_{sub} &=& \left|r_{01} +
\frac{t_{01}t_{10}\phi^2}{1-r_{10}r_{12}\phi^2}\right|^2\label{EqRsub}\\
R_{gr} &=& \left|\tilde{r}_{01} +
\frac{\tilde{t}_{01}\tilde{t}_{10}\phi^2}{1-\tilde{r}_{10}r_{12}\phi^2}\right|^2\label{EqRgr}
\end{eqnarray}

\noindent where indices 0,1 and 2 refer to vacuum ($\epsilon =
1$), SiO$_{2}$ and Si layers respectively and
\begin{eqnarray}
\phi =
\exp\left(i\frac{\omega}{c}\sqrt{\epsilon_{1}}d_{1}\right).
\end{eqnarray}

\noindent We used the complex reflection and transmission
coefficients at the interface between media $i$ and $j$:
\begin{eqnarray}
r_{ij}&=&\frac{\sqrt{\epsilon_{i}} -
\sqrt{\epsilon_{j}}}{\sqrt{\epsilon_{i}} +
\sqrt{\epsilon_{j}}}\\
t_{ij}&=&\frac{2\sqrt{\epsilon_{i}}}{\sqrt{\epsilon_{i}} +
\sqrt{\epsilon_{j}}}.
\end{eqnarray}

\noindent The presence of graphene between layers (in our case
it is between vacuum and SiO$_{2}$) modifies the interface
coefficients in the following way:
\begin{eqnarray}
\tilde{r}_{ij}&=&\frac{\sqrt{\epsilon_{i}} -
\sqrt{\epsilon_{j}}-\pi\alpha\frac{G}{G_0}}{\sqrt{\epsilon_{i}}
+ \sqrt{\epsilon_{j}}+\pi\alpha\frac{G}{G_0}}\\
\tilde{t}_{ij}&=&\frac{2\sqrt{\epsilon_{i}}}{\sqrt{\epsilon_{i}}
+ \sqrt{\epsilon_{j}}+\pi\alpha\frac{G}{G_0}}
\end{eqnarray}
\noindent where $\alpha=e^2/\hbar c$ is the fine structure
constant. The latter formulas are valid in the thin-film limit
(the thickness is much smaller than the wavelength), which is
perfectly applicable to graphene.

Since the typical values $\Delta R/R$ (Fig.\ref{FigReflS1}b)
are rather small ($\sim$ 10$^{-2}$), it is useful to introduce
the so-called "sensitivity" functions that we previously used
in similar analyses \cite{KuzmenkoPRL03,KuzmenkoPRB05} and
employ an approximate linear relation:

\begin{eqnarray}
\frac{\Delta R(\omega)}{R} \approx
\beta_{1}(\omega)\frac{\mbox{Re } \Delta G(\omega)}{G_{0}} +
\beta_{2}(\omega)\frac{\mbox{Im } \Delta G(\omega)}{G_{0}}
\end{eqnarray}

\begin{figure}[htp]
\includegraphics[width=8.5cm]{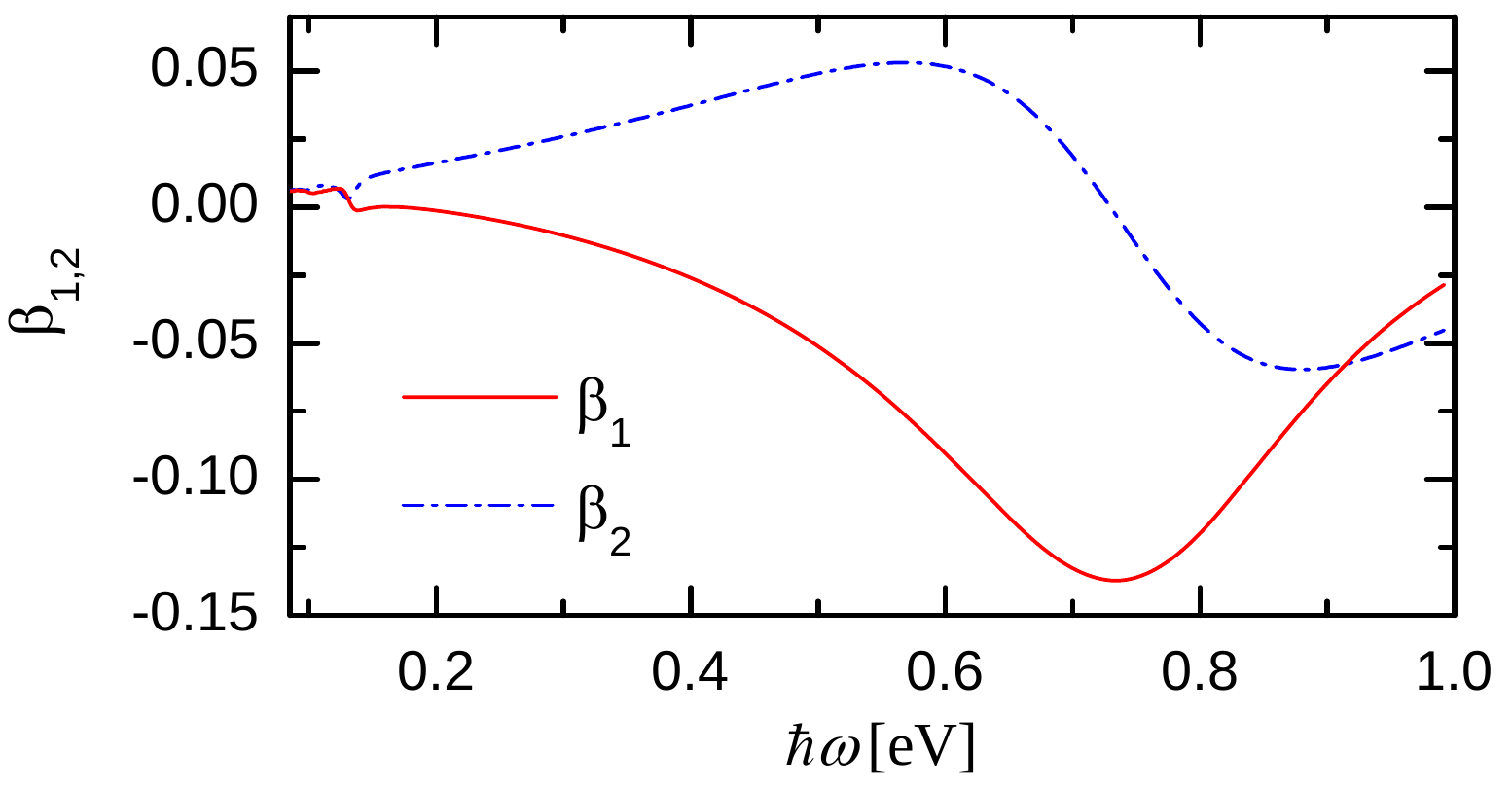}\\
\caption{(color on-line) The functions $\beta_{1}(\omega)$ and
$\beta_{2}(\omega)$, describing the sensitivity of $(\Delta
R/R)(\omega)$ to $\Delta G_{1}(\omega)$ and $\Delta
G_{2}(\omega)$ respectively (obtained at 10 K).}
\label{FigSensitivities}
\end{figure}

\noindent Here we obtained the sensitivity functions
$\beta_{1}(\omega)$ and $\beta_{2}(\omega)$ numerically, using
a linear regression of the exact formulas for the values of
$G/G_{0} \sim 1$. These function, which are specific to the
substrate used are shown in (Fig.\ref{FigSensitivities}). One
can see that the reflectivity depends on both the real and the
imaginary parts of $G(\omega)$ in a non-trivial way. At high
energies, the dependence is stronger than at low energies.
Optical phonons in SiO$_{2}$ give rise to structures at $\sim$
0.15 eV, which affect $\Delta R/R$.


\bibliography{biblio}

\end{document}